\documentclass[prl,twocolumn,aps,superscriptaddress]{revtex4}

\usepackage{graphicx}
\usepackage{bm}
\usepackage{amssymb}
\usepackage{amsmath}
\usepackage{subfigure}
\usepackage{tikz}

\begin{document}

\title{Kagome Lattice Hubbard model at half filling}
\author{Siegfried Guertler}
\affiliation{Lehrstuhl f\"ur Theoretische Physik II, Technische Universit\"at Dortmund, 44221 Dortmund, Germany}
\date{\today }

\begin{abstract}
We investigate the Kagome lattice Hubbard model at half-filling by variational Monte-Carlo with testing the U(1)-Dirac spin liquid, uniform and valence bond crystal states.
Even for the large-$U$ case the U(1) Dirac state, being the optimal state in the Heisenberg model, cannot be
recovered. While the finite $U$ Hubbard model allows the introduction of vacancies in a different manner compared
to the $t-J$ model, the physics appears to have many similarities. In particular a valence bond crystal is formed in the intermediate-$U$ regime. We observe an impact of the formation 
of this valence bond crystal on a possible Mott-transition in this model and discuss the properties of the wave-function.
\end{abstract}

\maketitle

{\it Introduction: } The Kagome lattice is the prototype lattice to study effects of geometric frustration. 
Numerical methods such as variational Monte-Carlo (VMC), density matrix renormalization group (DMRG), exact diagonalization (ED) and series expansion (SE) have been used to
clarify the ground state of the quantum Heisenberg model \cite{HUS1,iqb,white,NAK,LAEU,DEP} on the Kagome lattice. 
Initially various spin-liquid and valence bond crystal/solid states have been suggested, meanwhile there is strong indication for a spin-liquid state. The  classification and properties, including the questions whether it is a gapped or gapless spin-liquid, are still debated \cite{iqb,white,DEP,IQB2}. 
Recently interest in the doped and diluted variants of the model arose. In part because of ZnCu$_3$(OH)$_6$Cl$_2$, which is so far the best candidate of an compound realizing a structurally ideal Kagome lattice. This material consists of Kagome layers linked by Zn-ions. In the Kagome plane Copper ions form the Kagome-net and have an antiferromagnetic spin-1/2 interaction. During synthesis of this compound around five percent of the Zn-ions exchange their position with Cu-ions and, therefore, impurities are present \cite{Imai1,Mendels1,OLA,shores1,Vries1,Helton1}.

An earlier work by us found that within variational Monte-Carlo (VMC) the ground state of the $t-J$ changes 
drastically from a the Dirac spin-liquid at half filling to a valence bond crystal (VBC) \cite{GUERT1,GUERT2} for finite but small doping.
Another interesting subject is the Hubbard model, where even in the half-filled case double occupied and vacant sites are possible unless $U=\infty$.
The Hubbard model on the Kagome has been investigated with regard to a possible Mott-transitions \cite{MOTT1, MOTT2, MOTT3} and for the special case of van-Hove filling \cite{QIN,KI}. 
In actual compounds applying pressure would change the $t/J$ ratio, implying that experimentally one may reach a Hubbard model through this route. 
The investigation of the Hubbard model on the Kagome lattice provides an opportunity to study the effects of frustration for both, charge and spin-degree of freedom.

We address in this paper the physics of the half-filled Hubbard model on the Kagome lattice. We investigate if the Hubbard model for large $U$ is able to recover the U(1) Dirac spin-liquid state found in the Heisenberg model. We ask the question whether and how the physics differs when tuning $U$ in the Hubbard model, compared to introduction vacancies by doping in the infinite $U$ case \cite{GUERT1,GUERT2}, e.g. if a VBC state forms. Our paper is a first step to understand the complicated and possible rich physics in the Kagome lattice Hubbard model by a large scale numerical approach.\\

\begin{figure}
 \begin{center}
 \includegraphics[width=4.2cm]{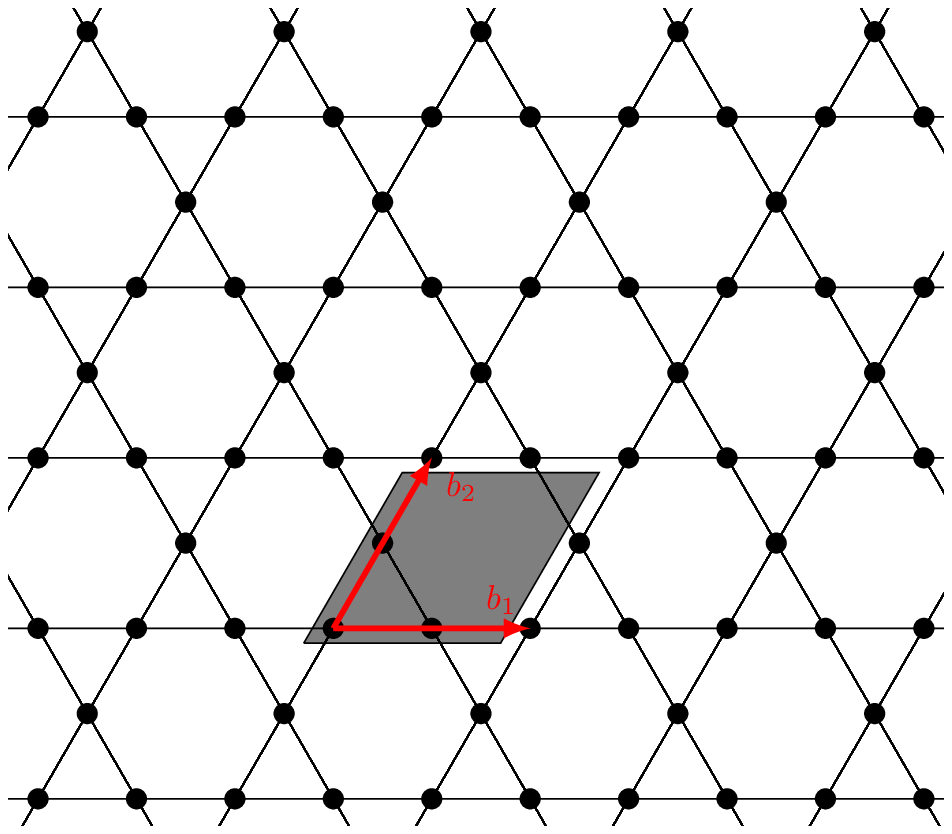}\\
 \vspace{5mm}
    \begin{tabular}{c|c}
    \resizebox{16mm}{!}{\includegraphics[]{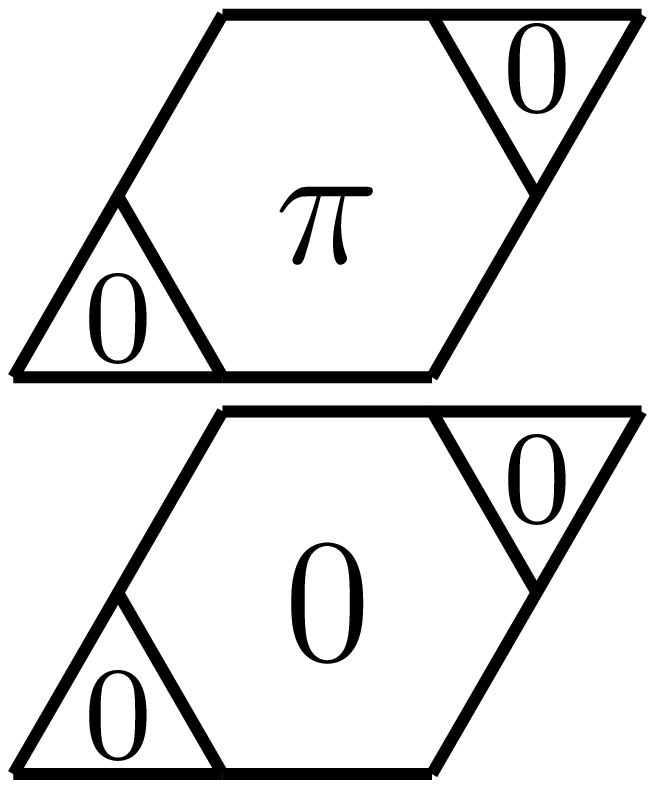}} &
    \resizebox{32mm}{!}{\includegraphics[]{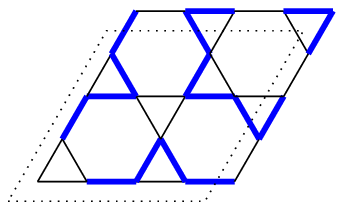}}
    \end{tabular}
\end{center}
 \caption{(Color online) Above: The Kagome lattice, consisting of corner-sharing triangles. Below: The variational states considered: Left the D-state (with $\pi$-flux in hexagons) and the U-state (with no flux); Right: HVBC-state; the blue thick bonds corresponds to the VBC of Hastings and are controlled by $\chi_1$.}
 \label{f1}
\end{figure}

{\it Model and Method: } As depicted in Fig. \ref{f1} the Kagome lattice consists of corner-sharing triangles and has a physical unit-cell of 3 sites. We study the standard Hubbard model on the Kagome lattice defined as:

\begin{eqnarray}
H=-t\sum_{\langle ij\rangle}\sum_{\sigma=\uparrow\downarrow}\left(c^{\dagger}_{i\sigma}c_{j\sigma}+ {\rm h.c.}\right)+U\sum_{i} n_{i\uparrow} \cdot n_{i\downarrow}
\label{ham}
\end{eqnarray}
where $c_{j\sigma}$ is the electron annihilation operator of an
electron with spin $\sigma$ on site $i$ and
$n_{i\sigma}=c_{i\sigma}^{\dagger} c_{i\sigma}$. The sum $\langle i,j\rangle$  is over all the n.n. pairs.
We set $t=1$ as the energy unit and the on-site repulsion $U$ as parameter.
Within VMC the best state in the Heisenberg model is the U(1) Dirac spin-liquid state \cite{HAS1,RAN1,IQB2}. This state has a $\pi$-flux in the hexagon, and zero-flux in the triangles. Competitive is the uniform spin-liquid state and
a VBC state with a 12-site unit cell proposed by Hastings \cite{HAS1} (see Fig. \ref{f1} for the 3 states). In the $t-J$-model the uniform spin-liquid state has a lower energy than the U(1) Dirac spin-liquid once we introduce holes by doping. Additionally the above mentioned VBC is formed for low doping \cite{GUERT1,GUERT2}. In the Hubbard model at half filling we introduce holes by allowing double-occupation when 
tuning $U$ away from the $U=\infty$ connecting the two models. We consider these 3 states in our investigation and denote them as ``D-state", ``U-state" and ``HVBC-state" in what follows. To capture the physics of the Hubbard model we implement two projectors: the partial Gutzwiller projection and the doublon-holon binding factors. The partial projector is defined as:

\begin{equation}
\mathcal{P}_{\alpha}= \prod_i  (1-\alpha n_{i\uparrow} n_{i\downarrow}),
\end{equation}
where the product is over all lattice sites. It projects out double occupied sites in dependence of the chosen variational parameter $\alpha$ with $\alpha=1.0$ meaning a full projection with no double occupied states. 
The doublon-holon binding factor is necessary for an accurate description of the intermediate to strong coupling region and is defined as:

\begin{eqnarray}
\mathcal{P}_q&=& \prod_i  (1-\mu_q \mathcal{Q}_j) \\
\mathcal{Q}_j&=& d_j \prod_{\vec{r}} (1-h_{j+\vec{r}})+h_j \prod_{\vec{r}}(1-d_{j+\vec{r}}),
\end{eqnarray}
with $d_j=n_{j\uparrow} n_{j\downarrow}$, $h_j=(1-n_{j\uparrow})(1-n_{j\downarrow})$ and $\mu_q$ being a variational parameter between 0 and 1.
Alternatively to this second projector one can include a $J$-term with spin-spin exchange in the Hamiltonian, as the so called ``Hubbard-Heisenberg"-model \cite{GUERTLER}. These projectors are acting on our base wave-functions:

\begin{eqnarray}
 |\Psi_{\text{HVBC},\alpha,q}\rangle&=&\mathcal{P}_q\mathcal{P}_{\alpha}  |\Psi_{\text{HVBC}}\rangle \\
 |\Psi_{\text{D},\alpha,q}\rangle&=&\mathcal{P}_q\mathcal{P}_{\alpha}  |\Psi_{\text{D}}\rangle \\
 |\Psi_{\text{U},\alpha,q}\rangle&=&\mathcal{P}_q\mathcal{P}_{\alpha}  |\Psi_{\text{U}}\rangle 
\end{eqnarray}

We use a standard VMC scheme with periodic boundary conditions. Earlier studies on the $t-J$ and Heisenberg model suggest the finite size effect to be tiny starting from sizes of $8 \times 8$ unit cells. We present data for the sizes of $8\times 8$ and $10 \times 10$ which show indeed little derivation from each other. Depending on the parameter-regime and its
convergence properties (e.g. larger $U$ requires longer runs) we use 8-64 independent runs per data point (defined by a specific set of parameters) for which 
we thermalize for 20.000-80.000 sweeps and measure for up to 700.000 sweeps.\\

\begin{figure}
   \begin{center}
    \includegraphics[width=0.94\columnwidth]{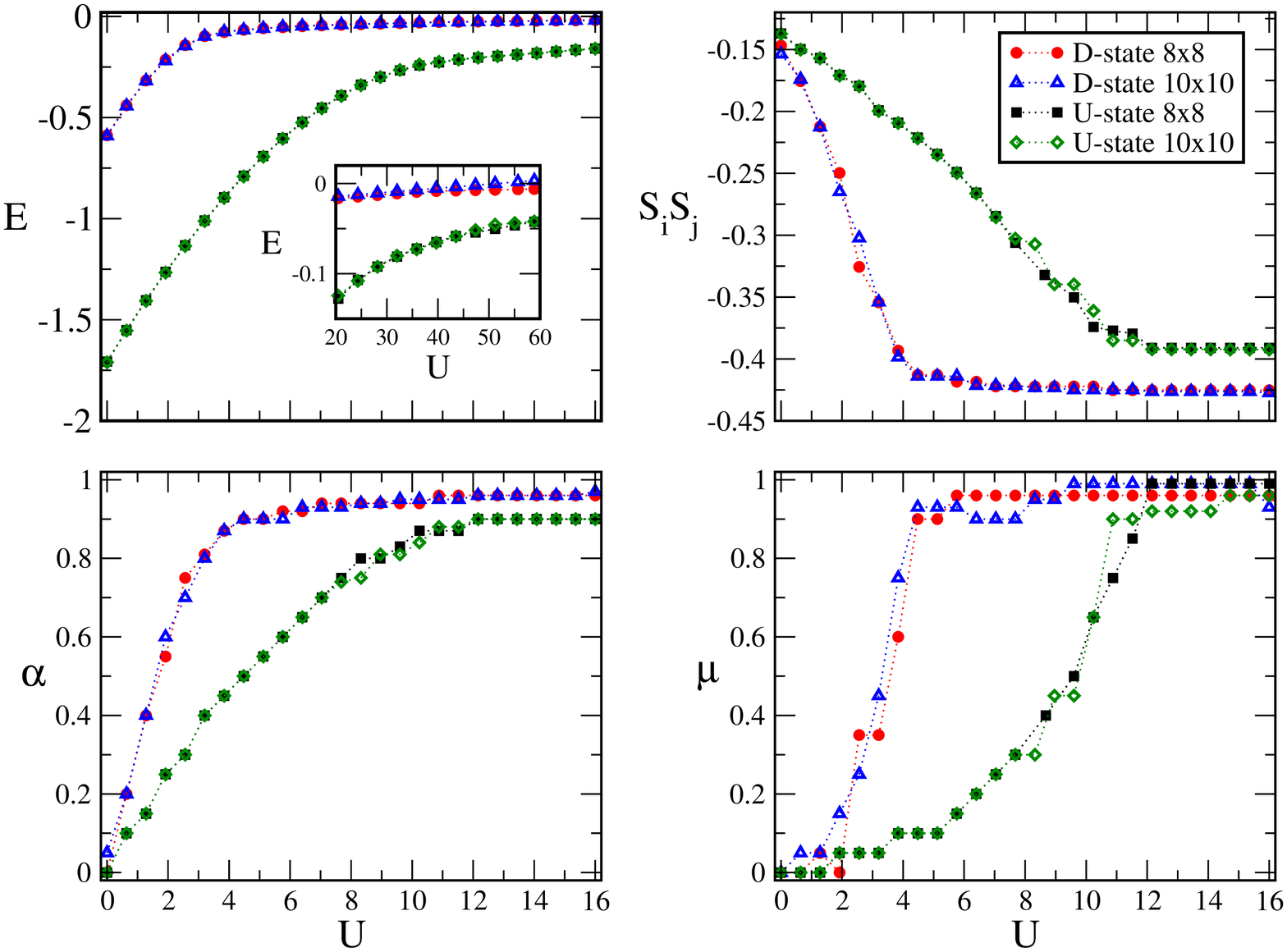}
   \end{center}
\caption{(Color online) Comparison between the U- and D-state for two sizes of the energy $E$ (a), spin-spin exchange $S_iS_j$ (b), projection $\alpha$ (c) and binding factor $\mu$ (d). The U-state has always lower energy.}
\label{UVSD}
\end{figure}

{\it Results: }  In Fig. \ref{UVSD} (a) we plot the energy per site of the U- and the D-state as a function of $U$. The energy of the D-state gradually comes closer to the one of the U-state, but it never has a lower energy (for values up to $U=60$, as can be seen in the inset of the same figure). The binding-factor $\mu$ rises sharply at $U\approx 11$ for the U-state and $U\approx 4$ for the D-state, indicating 
the position of a possible Mott-transition (Fig. \ref{UVSD} (c)).
We notice that both projectors $\alpha$ and $\mu$ rise quicker for the D-state than for the U-state. It appears that any introduction of vacancies and therefore mobility in the system renders the D-state as unfavourable. In Fig. \ref{UVSD} (b) we show the spin-spin exchange. After the Mott-transition this value reaches $S_iS_j\approx 0.40$ and saturates at around $S_iS_j\approx0.406$ for the U-state. The D-state reaches $S_iS_j\approx 0.428$ for large $U$ consistent with the result in the Heisenberg model. For the plots the two systemsizes are having almost the same values and there is not systematic shift in one direction for the values, thus the finite-size effect is tiny.

\begin{figure}
   \begin{center}
    \includegraphics[width=0.94\columnwidth]{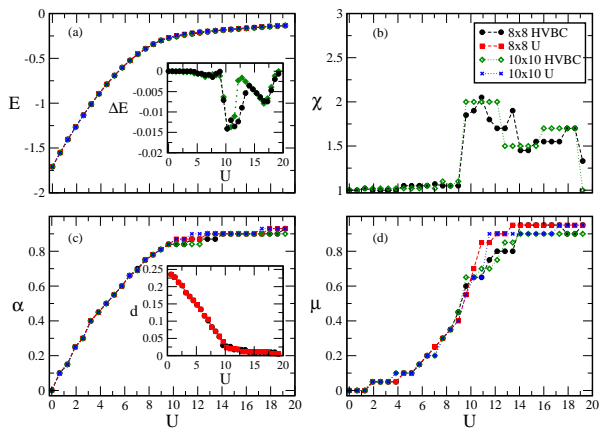}
   \end{center}
\caption{(Color online) Comparison of the HVBC-state with the U-state for two systemsizes.
(a) Optimized energy and in the inset the energy difference between the U- and HVBC state, (b) Variational parameter $\chi$, 
(c) projection parameter $\alpha$, in the inset measured double occupancy (shown only for the $8\times 8$ system) and (d) holon-binding factor $\mu$.}
\label{VBCDATA}
\end{figure}

\begin{figure}
   \begin{center}
    \includegraphics[width=0.75\columnwidth]{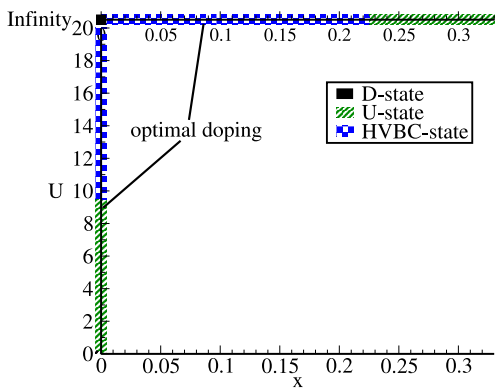}
   \end{center}
\caption{(Color online) Phase-diagram combined with the results of two earlier references by us. Indicated is as well the position of the optimal doping for the HVBC in the $t-J$ model, and the equivalent position in terms
of measured double-occupation in the Hubbard model.}
\label{TJHUBPHASE}
\end{figure}

Next we focus on the question whether a HVBC state is formed as it is the case for the $t-J$ model. The U-state is contained in the HVBC-state, as they are equivalent when $\chi=1.0$. We show all the relevant data (the variational parameters $\alpha$, $\mu$ and $\chi$ and the measured quantities energy $E$ and double occupancy $d$) in Fig. \ref{VBCDATA}. 
We compare the data of the U-state ($\chi=1.0$) with the one of the HVBC-state ($\chi$ varied) for two systemsizes. Note that the smaller size has a finer spacing in the parameters. In Fig. \ref{VBCDATA} (a) the energies and the energy difference (inset) are plotted. Starting from $U\approx9$ the HVBC state has a lower energy. The biggest energy-gain is at $U\approx10$ where $\Delta E\approx-0.015$ (error-bars at that position: $\approx0.002$ for $8 \times 8$ and $\approx0.008$ for $10 \times 10$). While for all $U > 9$ we find an energy gain for the HVBC-state, this gain is decreasing with larger $U$ and is within the error-bar for $U>18$. Observing the evolution of $\chi$ with $U$ we see a rather strong response of $\chi$ at $U=9$.
Note that we chose a fine spacing for $\chi$ for values close to $1.0$ ($\Delta \chi \approx 0.01$ for $8\times8$) and a larger spacing for the values close to $2.0$ ($\Delta \chi \approx 0.15$ for $8\times8$) capturing the overall situation well, and being a more economic solution for computation. There is little difference of the optimized value of $\alpha$ and the measured double occupation $d$ when comparing the HVBC-state with the U-state. The doublon-holon binding shows differences: For the U-state we see a sharp rise being a typical indicator of a Mott-transition (similar to studies of the unfrustrated Hubbard model with the same method), for the HVBC-state, $\mu$ takes longer to rise to the same value, and the area is smeared out. This area is exactly at the point where the HVBC state has the highest energy gain. Comparing the two sizes, we see that the agree very well, 
thus the finite-size effect appears not to play a role. 

\begin{figure}
   \begin{center}
    \includegraphics[width=0.85\columnwidth]{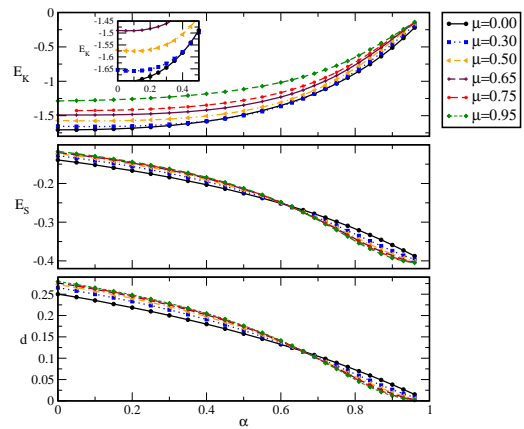} 
   \end{center}
\caption{(Color online) Impact of $\alpha$ on the energetics for given $\mu$: kinetic energy $E_K$, spin-spin exchange $E_S$ and double occupation $d$.}
\label{ENMUE}
\end{figure}

\begin{figure*}
   \begin{center}
    \begin{tabular}{c|c}
    \includegraphics[width=0.85\columnwidth]{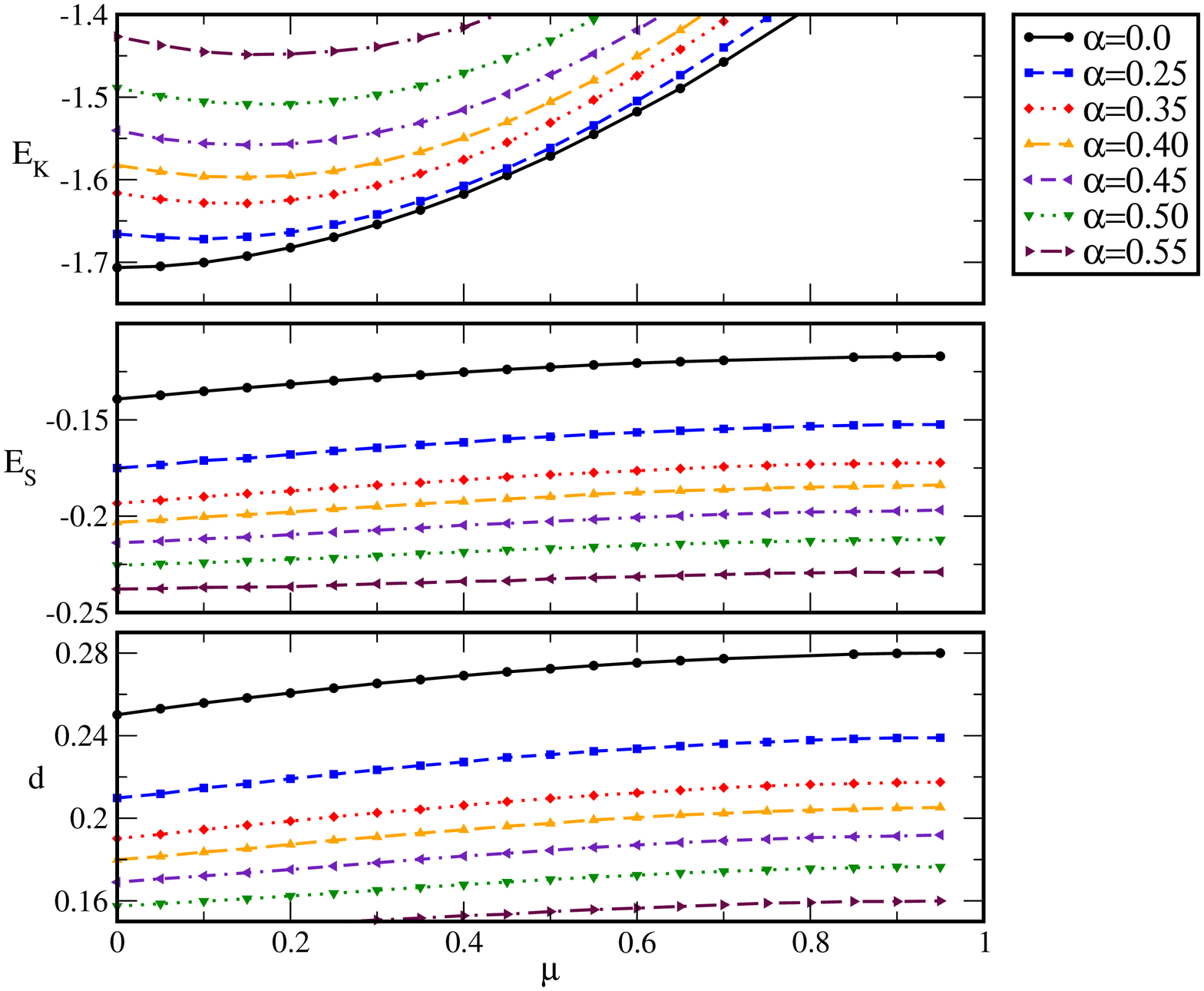} &
    \includegraphics[width=0.85\columnwidth]{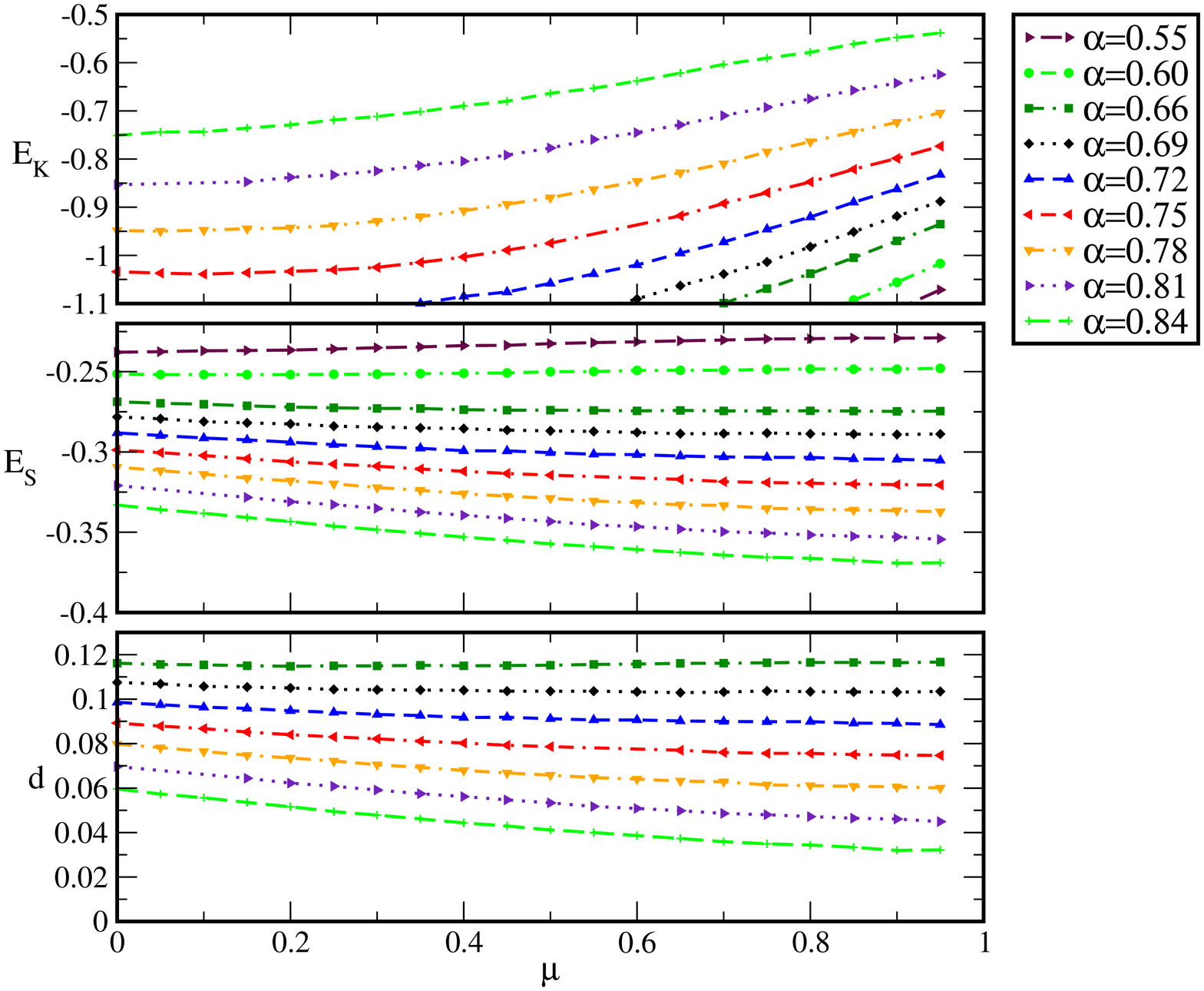}
    \end{tabular}
   \end{center}
\caption{(Color online) (left) Impact of $\mu$ on the energetics: kinetic energy $E_K$, spin-spin exchange $E_S$ and double occupation $d$ for low values of $\alpha$ (right) and high values of $\alpha$.}
\label{ENALPHA}
\end{figure*}

Comparing the introduction of holes in the $t-J$ model by doping and the impact of allowing double-occupation by reducing $U$, we realize that this gives a qualitatively similar result: 
Reduction of the value of $U$ introduces more holes and similar to the case of the $t-J$ model dimerization sets in, but is destroyed for a large amount of holes (here low $U$). 
The HVBC-state is found in the $t-J$-model at $0.05\lessapprox x\lessapprox 0.25$ \cite{GUERT1} and as argued by us in an earlier reference \cite{GUERT2} the optimal doping level for this state is at $x=\frac{1}{12}\approx0.083$. In the Hubbard model the hole-concentration is controlled by the projection operator and the doublon holon-factor (see discussion below). Measuring the double-occupation (Fig. \ref{VBCDATA} (c)) we find $\langle d \rangle=0.025$ at the point a the maximum of dimerization ($U=10.8$) and $\langle d \rangle=0.006$ at $U=18$ where the HVBC- and U-states differ so little that it is within the error-bar. Therefore fewer holes are present in the HVBC-state in the Hubbard model compared to the $t-J$ model. In Fig. \ref{TJHUBPHASE} we show the resulting phase-diagram for both models.

\begin{figure}
   \begin{center}
    \begin{tabular}{cc}
   \includegraphics[width=0.51\columnwidth]{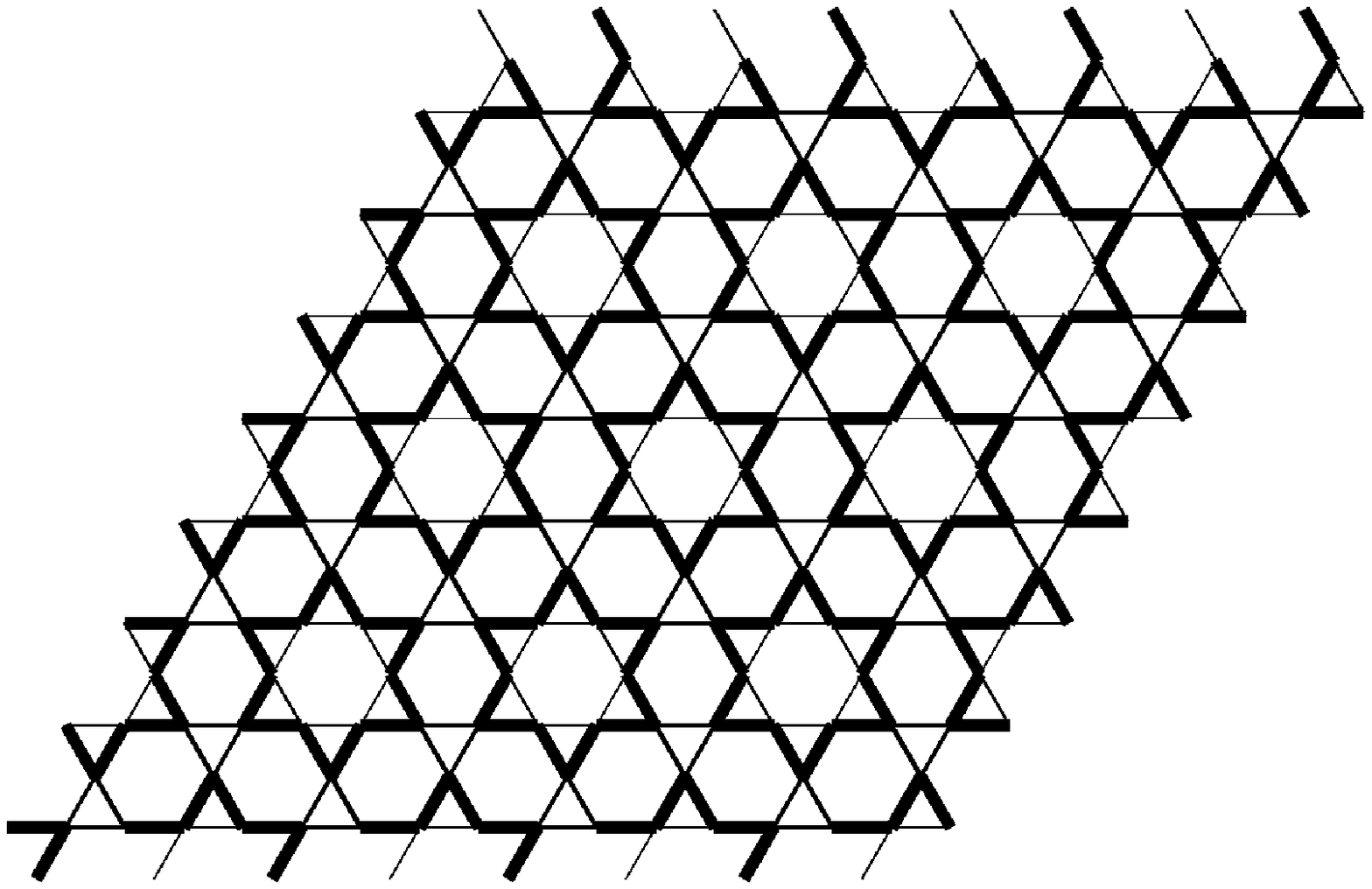} &
   \includegraphics[width=0.51\columnwidth]{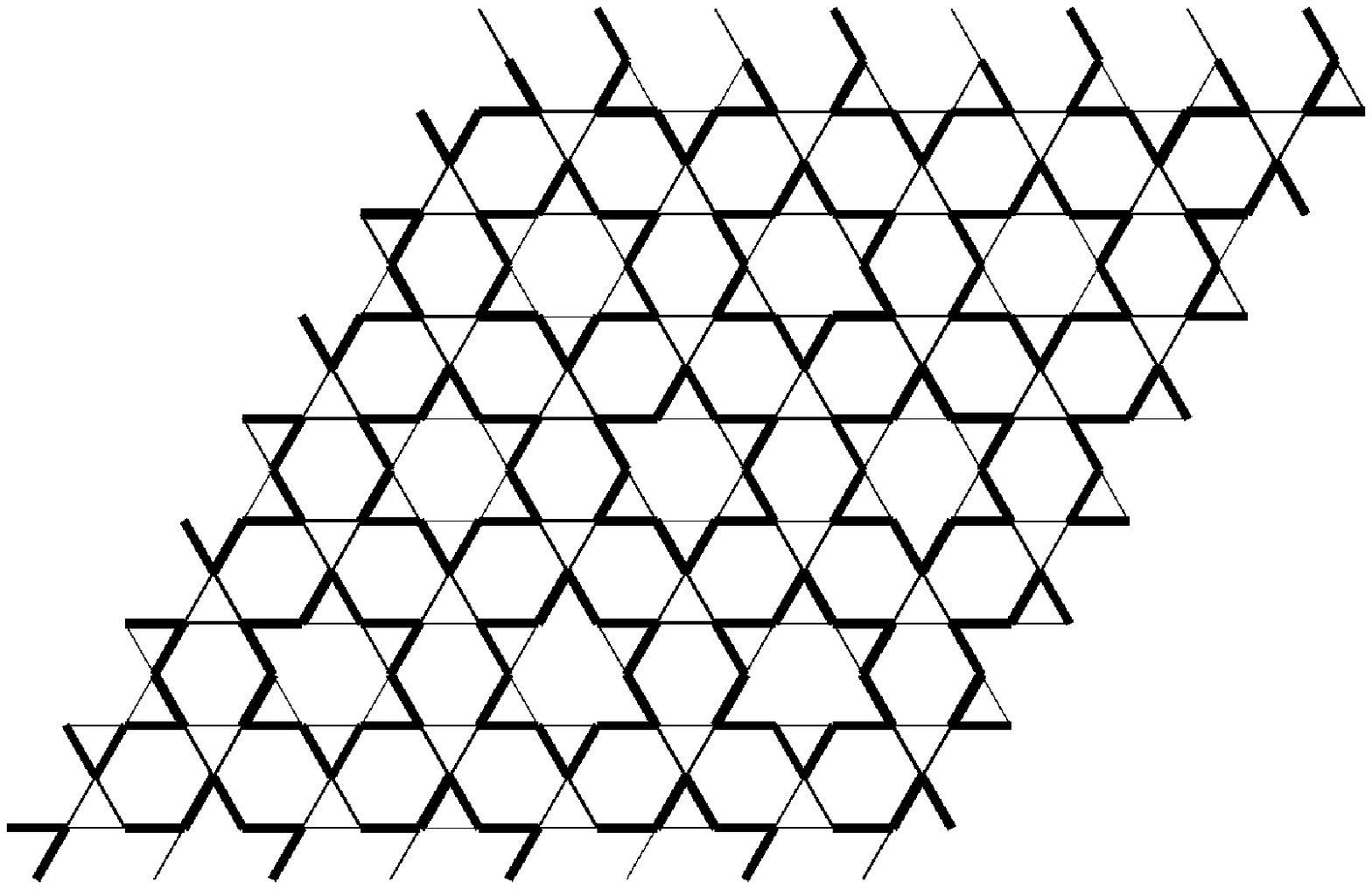}\\
   \includegraphics[width=0.51\columnwidth]{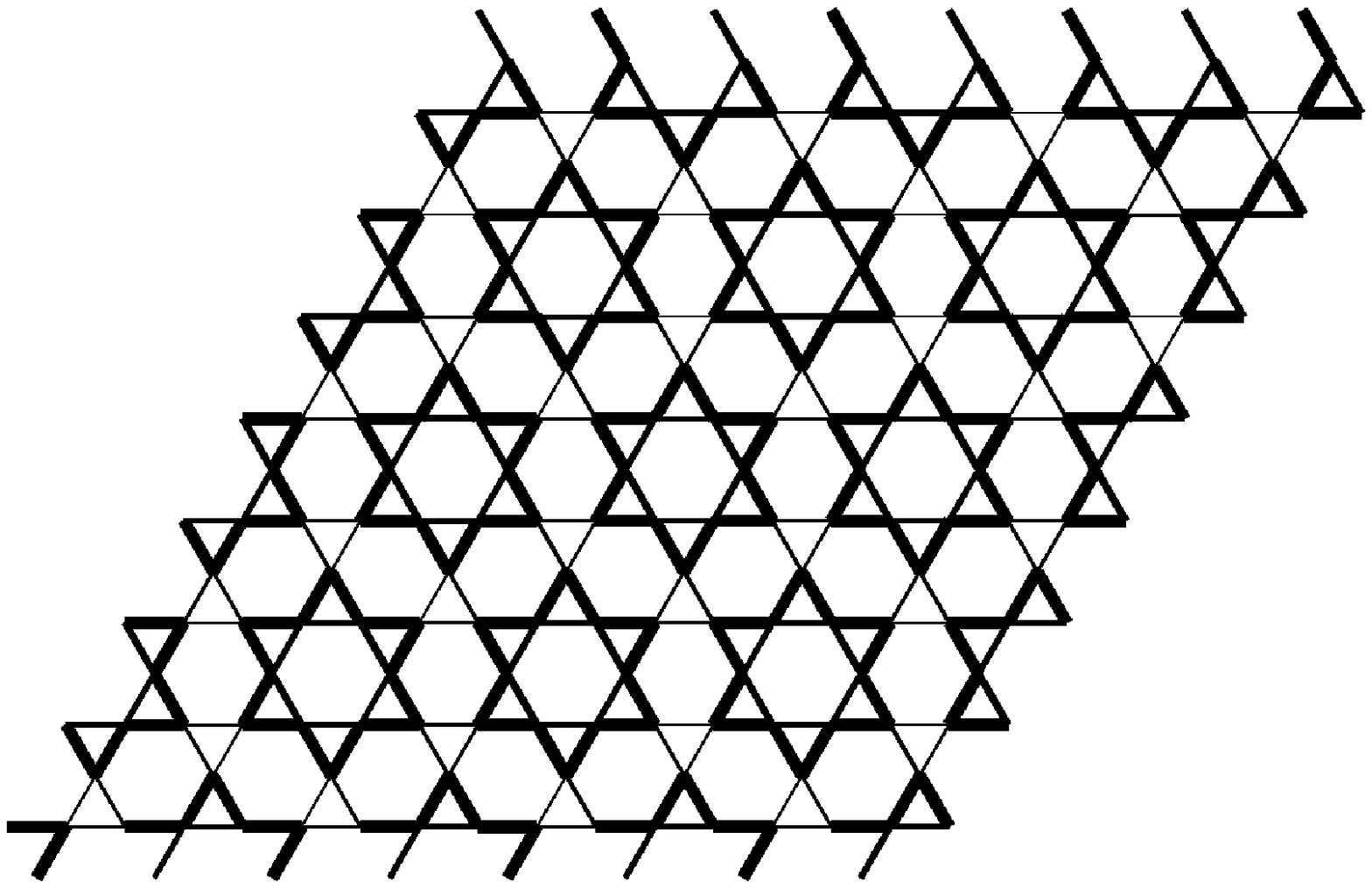} &
   \includegraphics[width=0.51\columnwidth]{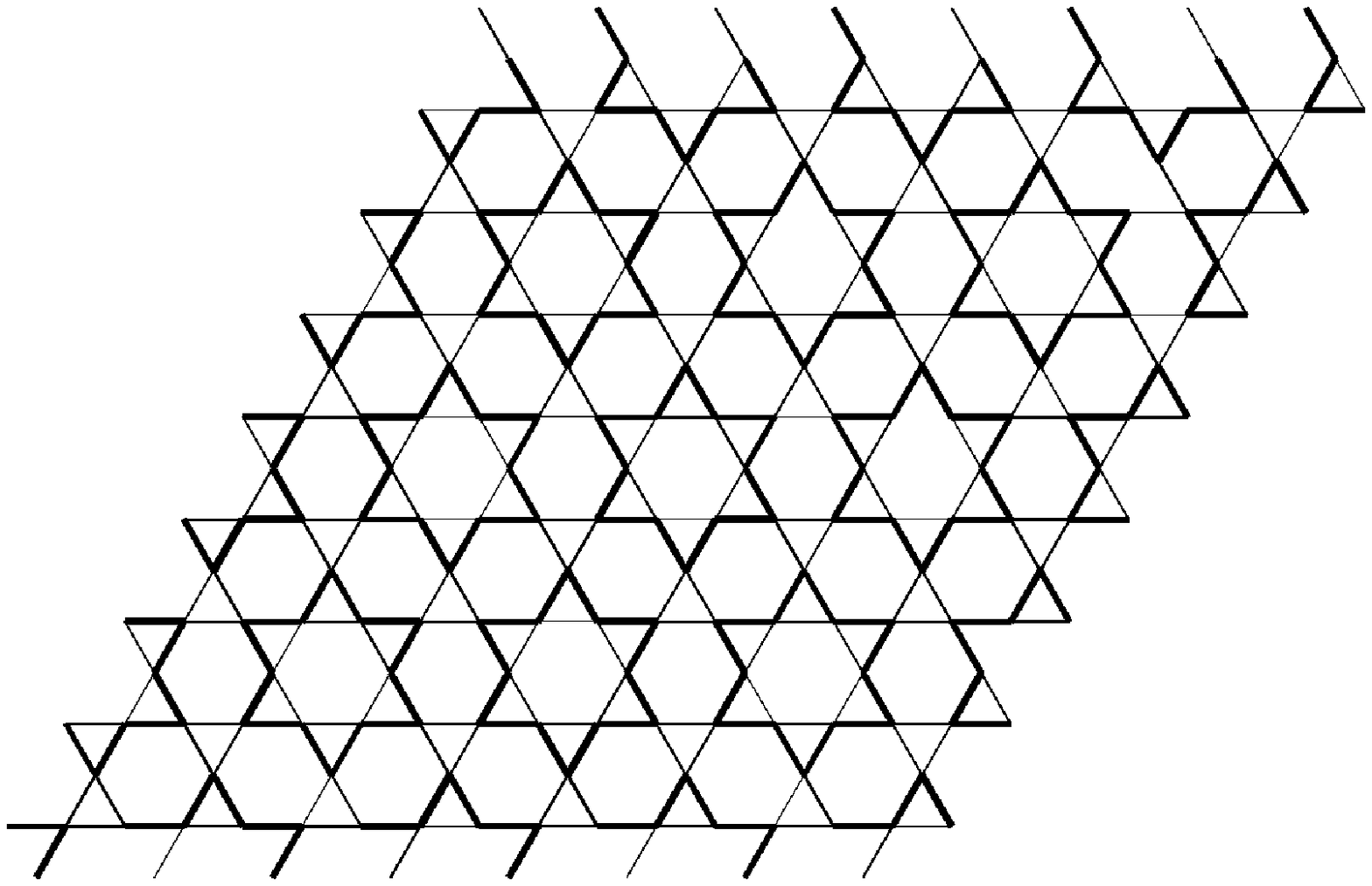}
    \end{tabular}
   \end{center}
\caption{Spatial variation of $E_K$ and $E_S$ for $\alpha=0.9$ and two values of $\mu$. Above: $E_S$. Below: $E_K$. Left: $\mu=0.0$ and Right: $\mu=0.99$.}
\label{SPINKINDENS}
\end{figure}

The wave-function for the Hubbard model has two additional parameters ($\alpha$ and $\mu$) compared to the wavefunction of the $t-J$ model. It is insightful to study the influence of them systematically: We investigate their impact on the kinetic energy $E_{K}=\langle c_i^{\dagger} c_j \rangle$, spin-spin exchange $E_{S}=\langle S_i S_j \rangle$ and double-occupation $n_d=\langle d \rangle$. We keep one of the projectors constant, vary the other one and measure the involved observables in the bulk and also locally on sites and bonds. We focus on the HVBC with $\chi \ne 1$. In Fig. \ref{ENMUE} we fix $\mu$ and vary $\alpha$. We start with $\mu=0$ and vary $\alpha$: $E_{S}$ will increase while $E_K$ is decreasing with increasing $\alpha$, as $\alpha$ controls $d$. For $0.0<\mu<0.7$ the combined effect is slightly more complicated as $E_{K}$ develops a minimum at $\alpha=0.05$ to $\alpha=0.10$ depending on the value of $\mu$. Fixing $\alpha$ and varying $\mu$ we
see two regimes which correspond to the phases below and above the Mott-transition (see Fig. \ref{ENALPHA}). For $0\le \alpha \le0.6$ $d$ is rising and therefore $E_{S}$ decreases, while for $\alpha>0.6$ there is the opposite trend. 

The local measurements of the observables reveals the spacial variation of them (see Fig. \ref{SPINKINDENS}). Large $\alpha$ and $\mu=0$ leads to the type of pattern found in the $t-J$ model at low doping: $E_{S}$ recovers the ``input" of the HVBC, $E_{K}$ develops a slight derivation of this pattern, having stronger contribution in the inner hexagon than on the triangles connecting the stars. 
For the case $\alpha=0.9$ and $\mu=0.99$ both $E_{S}$ and $E_{K}$ have a weak response to $\chi$, but the same pattern is found for both observables, showing that a large $\mu$ does not favour the previously proposed mechanism. Note that in the $t-J$ limit $\alpha=1.0$ while $\mu$ has no effect.

To summarize we investigated the Hubbard model at half-filling and addressed the question whether the D-state can be stabilized for any $U\ne\infty$, which appears to be not the case. This is similar to the $t-J$ model
where the D-state is well separated from the U-state. Thus the U-state will most likely be the more basic structure for the doped Hubbard model.
We study the formation of the HVBC-state which is qualitatively similar in both models. Quantitative differences between the two models can be attributed to the impact of the projection operators on the wave function. 
Another interesting finding is the impact of the HVBC on a possible Mott-transition. This might 
possibly be relevant for a larger class of frustrated models. We have summed up the combined result of this paper and the two preceding ones on the $t-J$ model in Fig. \ref{TJHUBPHASE}, where we have indicated the point of optimal doping in the $t-J$ model, and the corresponding point of the similar value for the double occupation in the Hubbard model. It is clear that doping the Hubbard model would show a much stronger response to this instability, which is an aspect we leave for future investigation. 

{\it Acknowledgments: } Supercomputer support was provided by the NIC, FZ J\"ulich under project
No. HDO07 and by ITMC of TU-Dortmund. Discussions with F. B. Anders and Q. H. Wang are acknowledged.


\begin{thebibliography}{50}

\bibitem{HUS1} R. R. P. Singh and D. A. Huse, Phys. Rev. B {\bf 76}, 180407 (2007).
\bibitem{iqb} Y. Iqbal, F. Becca, and D. Poilblanc, Phys. Rev. B, {\bf 84}, 020407(R) (2011).
\bibitem{white} S. Yan, D. A. Huse, and S. R. White, Science, {\bf 332} 1173 (2011).
\bibitem{NAK} H. Nakano and T. Sakai, J. Phys. Soc. Jpn. {\bf 80} 05370, (2011).
\bibitem{LAEU} A. M. Laeuchli, J. Sudan, and E. S. Sorensen, Phys. Rev. B {\bf 83}, 212401 (2011).
\bibitem{DEP} S. Depenbrock, I. P. McCulloch, and U. Schollwock, Phys. Rev. Lett., {\bf 109}, 067201 (2012).
\bibitem{IQB2}  Y. Iqbal, F. Becca, S. Sorella, and Didier Poilblanc, Phys. Rev. B, {\bf 87}, 060405 (2013).
\bibitem{Imai1} T. Imai, E. A. Nytko, B. M. Bartlett, M. P. Shores, and D. G. Nocera, Phys. Rev. Lett., {\bf 100}, 077203 (2008).
\bibitem{OLA} A. Olariu, P. Mendels, F. Bert, F. Duc, J. C. Trombe, M. A. de Vries, and A. Harrison, Phys. Rev. Lett., {\bf 100}, 087202 (2008).
\bibitem{shores1} M. P. Shores, E. A. Nytko, B. M. Bartlett, and D. G. Nocera, J. Am. Chem. Soc., {\bf 127}, 13462 (2005).
\bibitem{Mendels1} P. Mendels, F. Bert, M. A. de Vries, A. Olariu, A. Harrison, F. Duc, J. C. Trombe, J. S. Lord, A. Amato, and C. Baines, Phys. Rev. Lett., {\bf 98}, 077204 (2007).
\bibitem{Vries1} M. A. de Vries, K. V. Kamenev, W. A. Kockelmann, J. Sanchez-Benitez, and A. Harrison, Phys. Rev. Lett., {\bf 100}, 157205 (2008).
\bibitem{Helton1} J. S. Helton, K. Matan, M. P. Shores, E. A. Nytko, B. M. Bartlett, Y. Yoshida, Y. Takano, A. Suslov, Y. Qiu, J.-H. Chung, D. G. Nocera, and Y. S. Lee, Phys. Rev. Lett., {\bf 98}, 107204 (2007).
\bibitem{MOTT1} T. Ohashi, N. Kawakami, and H. Tsunetsugu, Phys. Rev. Lett., {\bf 97}, 066401 (2006).
\bibitem{MOTT2} S. Kuratani, A. Koga, and N. Kawakami, J. Phys.: Condens. Matter, {\bf 19}, 145252, (2007).
\bibitem{MOTT3} B. H. Bernhard, B. Canals, and C. Lacroix, J. Phys.: Condens. Matter, {\bf 19}, 145258, (2007).
\bibitem{QIN} W. S. Wang, Z. Z. Li, Y. Y. Xiang, Q. H. Wang, Phys. Rev. B 87, 115135 (2013).
\bibitem{KI} M. L. Kiesel, C. Platt, R. Thomale, Phys. Rev. Lett. 110, 126405 (2013).
\bibitem{GUERT1} S. Guertler and H. Monien, Phys. Rev. B, {\bf 84}, 174409 (2011).
\bibitem{GUERT2} S. Guertler and H. Monien, Phys. Rev. Lett., {\bf 111}, 097204 (2013).
\bibitem{HAS1} M. B. Hastings, Phys. Rev. B, {\bf 63}, 014413 (2000).
\bibitem{RAN1} Y. Ran, M. Hermele, P. A. Lee, and X. G. Wen, Phys. Rev. Lett., {\bf 98}, 117205 (2007).
\bibitem{GUERTLER} S. Guertler, Q. H. Wang, and F. C. Zhang, Phys. Rev. B, {\bf 79}, 144526 (2009).



\end{thebibliography}
\end{document}